\documentclass[10pt,twocolumn,english]{article}
\usepackage{lmodern}

\usepackage[T1]{fontenc}
\usepackage[latin9]{inputenc}
\usepackage{verbatim}
\usepackage{amsmath}
\usepackage{amssymb}

\makeatletter
\providecommand*{\code}[1]{\texttt{#1}}

\@ifundefined{date}{}{\date{}}
\makeatother

\usepackage{babel}
\begin{document}
\title{Java Generics: An Order-Theoretic Approach\\
(Abridged Outline)\thanks{A less-structured, but more-detailed outline of the order-theoretic
approach to modeling generics is presented in~\cite{AbdelGawad2019g}.
The two outline papers, together with~\cite{AbdelGawad2019h} which
outlines the use of category theory in the approach, succinctly summarize
the main points of the detailed articles~\cite{AbdelGawad2016a,AbdelGawad2016c,AbdelGawad2017a,AbdelGawad2017b,AbdelGawad2018a,AbdelGawad2018b,AbdelGawad2018c,AbdelGawad2018e,AbdelGawad2019a,AbdelGawad2019b,AbdelGawad2019}.
As motivating and illustrating as examples may be, to shorten the
three outline papers we intentionally elide most code examples as
well as examples of subclassing and subtyping relations that illustrate
the construction of the generic subtyping relation and other features
of the approach. To aid readers interested in examples or more details
however, the outline articles always cite one detailed article (or
more) while discussing each piece of the approach.}}
\author{Moez A. AbdelGawad\\
Informatics Research Institute, SRTA-City, Alexandria, Egypt\\
\texttt{moez@cs.rice.edu}}
\maketitle
\begin{abstract}
The mathematical modeling of generics in Java and other similar nominally-typed
object-oriented programming languages is a challenge. In this short
paper we present the outline of a novel order-theoretic approach to
modeling generics, in which we also elementarily use some concepts
and tools from category theory. We believe a combined order-theoretic
and category-theoretic approach to modeling generics holds the keys
to overcoming much of the adversity found when analyzing features
of generic OO type systems.
\end{abstract}

\section{Introduction}

Generics have been added to Java so as to increase the expressiveness
of its type system~\cite{JLS05,JLS18,Bracha98,Corky98,Thorup99}.
Generics in Java and other mainstream nominally-typed OOP languages
similar to it such as C\#~\cite{CSharp2015}, Scala~\cite{Odersky14},
C++~\cite{CPP2017}, and Kotlin~\cite{Kotlin18}, however, include
some features---such as variance annotations (\emph{e.g.}, Java wildcards),
$F$-bounded generics, and Java erasure---that have been hard to
analyze and reason about so far~\cite{Torgersen2004,MadsTorgersen2005,Cameron2007,Cameron2008,Summers2010,Tate2011}.
As a result, the type systems of mainstream nominally-typed OOP languages,
which are built based on current mathematical models of generics,
are overly complex\footnote{Check, for example,~\cite{GenericsFAQWebsite}, or sections of the
Java language specification that specify crucial parts of its generic
type system, \emph{e.g.}, \cite[$\mathsection$4.5 \& $\mathsection$5.1.10]{JLS18}.}, thereby hindering the progress of these type systems.

Further, support of some features in Java generics has a number of
irregularities or ``rough edges.'' These include type variables
that can have upper $F$-bounds but cannot have lower bounds (let
alone lower $F$-bounds), wildcard type arguments that can have an
upper-bound or a lower-bound but not both, and Java erasure---a feature
prominent in Java and Java-based OOP languages such as Scala and Kotlin---that
is usually understood, basically, as being ``outside the type system.''

In this short paper we outline a new order-theoretic approach to modeling
Java generics, and we report on our progress in developing this approach.
The main contribution of the approach is demonstrating how concepts
and tools from order theory can significantly simplify analyzing and
reasoning about subtle features of OO generics, including the ones
mentioned above. 

Fundamentally, in the approach we use the nominal \emph{subclassing}
relation (as a partial ordering between classes\footnote{In this work Java interfaces and Scala traits are treated as abstract
classes. In this paper the term `class' thus refers to classes and
other similar type-constructing constructs. Also, in other OOP literature
parameterized types are sometimes called \emph{object types}, \emph{class
types}, \emph{reference types}, \emph{generic types},\emph{ }or just
\emph{types}.}) together with some novel order-theoretic tools to construct the
generic nominal \emph{subtyping} relation (also a partial ordering,
between parameterized types) and the \emph{containment} relation (a
third partial ordering, between generic type arguments). These three
ordering relations lie at the heart of mainstream generic OO type
systems. Using order theoretic tools, as well as some concepts and
tools from category theory, we further analyze these three relations
and the relationship between them. Consequently, we demonstrate the
value of the approach by exploring extensions of generic OO type systems
that are naturally suggested by such analysis.

\section{Description}

\paragraph{Constructing The Generic Subtyping Relation}

The first step in the order-theoretic approach to modeling generics
is defining two operators that construct ordering relations (\emph{i.e.},
posets)%
. In particular, the first operator, called \emph{ppp} and denoted
$\ltimes$, takes two input posets and a subset of the first poset
and constructs a \emph{partial poset product}~\cite{AbdelGawad2018a,Davey2002}
of the two input posets. The second operator, called $wc$ (for \emph{wildcards})
and denoted $\triangle$, takes as input a bounded poset (\emph{i.e.},
one with top and bottom elements) and constructs a ``triangle-shaped''
poset---corresponding to wildcard type arguments---that is roughly
composed of three copies of the input poset. See~\cite{AbdelGawad2018b}
for more details on $\ltimes$ and $\triangle$.

The formal definition of \emph{ppp }is the order-theoretic counterpart
of the definition of partial Cartesian product of graphs presented
in~\cite{AbdelGawad2018a,AbdelGawad2018b}, while the formal definition
of the wildcards operator \emph{wc} is presented in~\cite{AbdelGawad2018b}.
It is worthy to note that if the input poset of \emph{wc} is a chain
(\emph{i.e.}, a ``straight edge''), then $wc$ will construct an
exact triangle-shaped output poset. The poset constructed by \emph{wc
}is ``triangle-shaped'' due to the existence of three variant subtyping
rules for generic wildcard types in generic OOP (where covariant subtyping,
roughly, is modeled by the left side of the triangle, contravariant
subtyping is modeled by the right side, while the base of the triangle
models invariant subtyping). See~\cite{AbdelGawad2018b,AbdelGawad2017a}
for details and examples.

Next, given a finite subclassing relation $C$,\footnote{In $C$ it is assumed that a generic class takes one type argument,
and that if a generic class extends (\emph{i.e.}, inherits from, or,
is a subclass of) another generic class then the superclass is passed
the parameter of the subclass as the superclass type argument (\emph{e.g.},
as in the declaration \code{\textbf{class~}D<T>\textbf{~extends~}C<T>},
where \code{T}, the type parameter of \code{D}, is used ``as is''
as the type argument of superclass \code{C}). While we do not expect
any significant complications when these simplifying assumptions are
relaxed, we keep a discussion of how these assumptions can be relaxed
to future work.

It is worthy to mention that the second assumption, in some sense,
models the most general case (of a type argument passed to the superclass)
and that a more complex inheritance relation (such as \code{\textbf{class~}D<T>\textbf{~extends~}C<E<T>\textcompwordmark >})
only restricts the set of \emph{valid} subtyping relations between
instantiations of the subclass (\emph{e.g.}, \code{D}) and those
of its superclasses (\emph{e.g.}, \code{C}). (See later discussion
of \emph{valid} versus \emph{admittable} subtyping relations).} operators \emph{ppp} and \emph{wc} are used to construct, iteratively,
the infinite subtyping relation $S$ between ground parameterized
types\footnote{Ground parameterized types are ones with no type variables. Such types
are infinite in number due to the possibility of having arbitrary-depth
nesting of type arguments~\cite{AbdelGawad2017a}. Subtyping between
these types is the basis for defining the full subtyping relation
that includes type variables~\cite{FJ/FGJ,TAPL}.}. In particular, given $S_{i}$, a finite approximation of $S$, \emph{wc}
constructs the corresponding wildcard type arguments, ordered by containment.
Given $C$ and the constructed arguments, \emph{ppp} then%
{} pairs generic classes %
with these arguments and adds types corresponding to non-generic classes%
{} to construct the poset $S_{i+1}$, %
ordered by subtyping, that next approximates $S$.\footnote{This iterative construction process constructs the (least fixed point)
solution of the recursive poset equation
\[
S=C\ltimes_{C_{g}}\triangle\left(S\right)
\]
where $S$ is the subtyping relation, $C$ is the subclassing/inheritance
relation, and $C_{g}$ is the subset of generic classes in $C$. See~\cite{AbdelGawad2018b}
for more details.}

\paragraph{\label{sec:The-Erasure-Connection}The Erasure Galois Connection
and Nominal Subtyping}

Erasure---where, intuitively, type arguments of a parameterized type
are ``erased''---is a feature prominent in Java and Java-based
OO languages such as Scala and Kotlin, but that can be also defined
and made explicit in other generic nominally-typed OOP languages.
In the order-theoretic approach to modeling generics, erasure is modeled
as a mapping from types to classes.\footnote{To model \emph{erased} \emph{types}, the erasure mapping is composed
with a notion of a \emph{default type} that maps each generic class
to some corresponding parameterized type (\emph{i.e.}, a particular
instantiation of the class). We keep further discussion of \emph{default
type arguments} and \emph{default types} to future work.}

Also, in the order-theoretic approach to modeling generics the `most
general wildcard instantiation' of a generic class is called the
\emph{free type} corresponding to the class. For example, a generic
class \code{C} with one type parameter has the type \code{C<?>}
as its corresponding free type\footnote{A \emph{non-}generic class is mapped to the only type it constructs---a
type typically homonymous to the class---as its corresponding free
type.}.

By maintaining a clear separation between classes ordered by subclassing,
on one hand, and types ordered by subtyping, on the other, the construction
of the subtyping relation using the subclassing relation (as presented
earlier, using order-theoretic tools) allows us to observe that a
Galois connection~\cite{Davey2002} exists between the two fundamental
relations in generic nominally-typed OOP. This connection between
subclassing and subtyping is expressed formally using the erasure
and free type mappings.

In particular, if $E$ is the erasure mapping that maps a parameterized
type to the class used to construct the type, and if $FT$ is the
free type mapping that maps a class to its most general wildcard instantiation,
then the connection between subclassing and subtyping states that
for all parameterized types $t$ and classes $c$ we have 
\begin{equation}
E\left(t\right)\leq c\Longleftrightarrow t<:FT\left(c\right)\label{eq:EGC}
\end{equation}
where $\leq$ is the subclassing relation between classes and $<:$
is the subtyping relation between parameterized types.\footnote{For example, in Java the statement 
\[
\mathtt{LinkedList\le List\Longleftrightarrow LinkedList\negthickspace<\negthickspace String\negthickspace>\;<:\;List\negthickspace<?\negthickspace>},
\]
where $t$, in Equation~(\ref{eq:EGC}) on page~\pageref{eq:EGC},
is instantiated to type \code{LinkedList<String>} and $c$ is instantiated
to class \code{List}, asserts that class \code{LinkedList} being
a subclass of \code{List} is equivalent to (\emph{i.e.}, if and only
if, or implies and is implied by) \code{LinkedList<String>} being
a subtype of the free type \code{List<?>}, which is a true statement
in Java.}\textsuperscript{,}\footnote{Based on the strong relation between category theory and order theory---see
below---the Galois connection between subclassing and subtyping is
called \emph{JEA},\emph{ }the Java erasure adjunction.}

It should be noted that the erasure Galois connection expresses a
fundamental property of generic nominally-typed OOP, namely, that
subclassing (\emph{a.k.a.}, type inheritance) is \emph{the} source
of subtyping in generic nominally-typed OOP. In other words, the property
states, in one direction, that type inheritance is \emph{a} source
of subtyping (\emph{i.e.}, subclassing causes subtyping between parameterized
types) and, in the other direction, that type inheritance is the\emph{
only} source of subtyping in generic nominally-typed OOP (\emph{i.e.},
subtyping between parameterized types comes from nowhere else other
than subclassing). This property of generic nominally-typed OOP---stated
as `inheritance is \emph{the }source of subtyping'---corresponds
to the `inheritance \emph{is }subtyping' property of \emph{non}-generic
nominally-typed OOP~\cite{NOOPsumm,InhSubtyNWPT13}.

\paragraph{Extending Generics: Interval Types and Doubly $F$-bound\-ed Generics}

The construction of the generic subtyping relation using tools from
order theory suggests how generics in nominally-typed OOP languages
can be extended in two specific directions.

First, the approach suggests how wildcard type arguments can simultaneously
have lower bounds and upper bounds, thereby defining \emph{interval
type arguments} (as a generalization of wildcard type arguments) and
defining \emph{interval types} (as a generalization of wildcard types---which
are parameterized types with wildcard type arguments). In particular,
interval types and the subtyping relation between them can simply
be constructed by replacing the \emph{wc }operator (that we presented
earlier) with a novel operator \emph{int} ($\Updownarrow$) that constructs
interval type arguments (and the containment relation between them)
from an input subtyping relation.\footnote{The formal definition of operator \emph{int }is presented in~\cite{AbdelGawad2018c}.
Unlike the \emph{wc }operator, operator \emph{int }does not require
the input poset to be bounded, \emph{i.e.}, it does not assume the
existence of a greatest type \code{Object} and a least type \code{Null}
in the input (subtyping) relation.} The subtyping relation is then iteratively constructed from the subclassing
relation, again as the solution of a recursive poset equation.\footnote{Namely, the equation $S=C\ltimes_{C_{g}}\Updownarrow\left(S\right)$.
See~\cite{AbdelGawad2018c} for more details.}

Second, the approach suggests how to define \emph{doubly $F$-bounded
generics} (\emph{dfbg}, for short),\emph{ }where a type variable can
have both an upper $F$-bound and a lower $F$-bound.\footnote{It is worthy to note that the definition of \emph{dfbg} got inspiration
from functions in real analysis. See~\cite{AbdelGawad2018e} for
more details.} See~\cite{AbdelGawad2018e} for more details on \emph{dfbg}.

Considering \emph{dfbg }led us to distinguish between \emph{valid
type arguments}, which satisfy declared type parameter bounds, and
\emph{admittable type arguments}, which do not necessarily satisfy
bounds, and to thus define \emph{valid parameterized types}, whose
type arguments are valid, and \emph{admittable parameterized types},
whose type arguments are admittable but may not be valid (such as
the admittable but invalid Java type \code{Enum<Object>}), and, accordingly,
to define \emph{valid subtyping relations}, between valid parameterized
types, and \emph{admittable subtyping relations}, between two admittable
parameterized types that are not necessarily valid.%

\paragraph{Induction, Coinduc\-tion, (Co)Ind\-uctive Types and Mutual (Co)Ind\-uction
in Generic OOP}

In logic, \emph{coinductive} reasoning can, intuitively, be summarized
as asserting that a statement is proven to be true if there is \emph{no}
(finite or ``good'') reason for the statement to \emph{not} hold~\cite{Kozen2016,AbdelGawad2019a}.
While analyzing \emph{dfbg }in~\cite{AbdelGawad2018e}, we use a
coinductive logical argument to prove that checking the validity of
type arguments inside some particular bounds-declarations of generic
classes is unnecessary. Also, in~\cite{Tate2011} Tate et al. conclude
that Java wildcards are some form of coinductive\emph{ }bounded existentials.\footnote{Given their historical origins~\cite{Knaster1928,Tarski1955}, induction
and coinduction---and accordingly (co)inductive mathematical objects---are
naturally best studied in lattice theory, which is a sub-field of
order theory.}

Combined, these factors motivated us to consider, in some depth, the
status of (co)inductive types in our order-theoretic approach~\cite{AbdelGawad2019b},
which led us to define the notions of $F$-subtypes and $F$-supertypes
of a generic class $F$.\footnote{A parameterized type \code{Ty} is an $F$-subtype of class $F$ iff
\code{Ty~<:~F<Ty>}. Dually, type \code{Ty} is an $F$-supertype
of class $F$ iff \code{F<Ty>~<:~Ty}. The names of these two notions
come from category theory (see later discussion), where $F$-subtypes
correspond to $F$-coalgebras while $F$-supertypes correspond to
$F$-algebras of a generic class $F$ (called a functor $F$ in category
theory, and a generator---or a constructor---$F$ in lattice theory
and order theory).} The value of defining these notions is illustrated in the definition
of \emph{dfbg}, where a type variable, say \code{T}, with both a
lower $F$-bound and an upper $F$-bound ranges over a set of $F$-supertypes
and $F$-subtypes specified by the bounds of \code{T}. See~\cite{AbdelGawad2018e,AbdelGawad2019b}
for more details.

Further, the mutual dependency between the containment relation (as
an ordering of generic type arguments) and the subtyping relation
(as an ordering of parameterized types), in addition to the fact that
classes in OO programs, including generic classes, are frequently
defined mu\-tually-recur\-sively\footnote{\emph{E.g.}, assuming the absence of primitive types in Java, the
definitions of classes \code{Object} and \code{Boolean} are mutually
dependent on each other (since class \code{Boolean} extends \code{Object},
and, without primitive type \code{bool}, the fundamental \code{equals()}
method in \code{Object} returns a \code{Boolean}).}, led us to also define an order-theoretic notion of \emph{mutual
}(\emph{co})\emph{induction} to allow studying least and greatest
fixed point solutions of mutually-recursive definitions\footnote{Which are common in OOP but also in programming in general.}
in an order-theoretic context~\cite{AbdelGawad2019}.

\paragraph{\label{sec:Category-Theory}Using Category Theory in Modeling Generics}

Category theory can be viewed simply as a (major) generalization of
order theory~\cite{Fong2018,Priestley2002,spivak2014category}. In
particular, each poset can be viewed, canonically, as a (\emph{thin},
\emph{small})\emph{ }category~\cite{Fong2018}.

As such, some concepts and tools from category theory, such as adjunctions,
monads, $F$-(co)algebras, initial algebras (\emph{e.g.}, co-free
types), final coalgebras (\emph{e.g.}, free types), and operads, can
be used to generalize the order-theoretic model of generics, and to
situate it in the context of category theory.

A more detailed account of the use of category theory in our approach
is presented in~\cite{AbdelGawad2019h}.

\section{Discussion}

In this short paper we presented the outline of an order-theoretic
model of generic nominally-typed OOP. This model demonstrates that
in generic nominally-typed OOP:\vspace{0.05cm}

\noindent \quad{}$\bullet$\enskip{}The subtyping relation between
parameterized types can be constructed solely from the subclassing
(\emph{i.e.}, inheritance) relation between classes using order-theoretic
tools,

\noindent \quad{}$\bullet$\enskip{}Erasure can be modeled as a
map (\emph{i.e., }a homomorphism) from parameterized types ordered
by subtyping to classes ordered by subclassing,

\noindent \quad{}$\bullet$\enskip{}Wildcard type arguments can
be modeled as intervals over the subtyping relation,\footnote{\noindent In particular, intervals with upper bound \code{Object}
or lower bound \code{Null}.}

\noindent \quad{}$\bullet$\enskip{}Generic classes can be modeled
as type generators over the subtyping relation\footnote{\noindent I.e., as mathematical functions that take in type arguments,
ordered by containment, and produce parameterized types, ordered by
subtyping.},

\noindent \quad{}$\bullet$\enskip{}The complex \code{\noindent open}
and \code{\noindent close} operations (\emph{i.e.}, capture conversion;
see~\cite[$\mathsection$5.1.10, p.113]{JLS18}) are \emph{not} needed
in the definition of the subtyping relation between ground parameterized
types\footnote{\noindent Ground parameterized types constitute the full set over
which type variables of generic classes range.}, since the relation can be defined exclusively using the containment
relation between generic type arguments, and

\noindent \quad{}$\bullet$\enskip{}Upper $F$-bounded type variables---\emph{e.g.},
type variable \code{T} in the class declaration \code{\textbf{class }Enum<T \textbf{extends} Enum<T>\textcompwordmark >}---range
over $F$-subtypes (modeled as coinductive types of the type generators
modeling generic classes), while lower $F$-bounded type variables
range over $F$-supertypes (modeled as inductive types).\vspace{0.05cm}

Moreover, the model hints that:\vspace{0.05cm}

\noindent \quad{}$\bullet$\enskip{}Circular (\emph{a.k.a.}, infinite,
or infinitely-justified) subtyping relations can be modeled by an
order-theoretic coinductive interpretation~\cite{KennedyDecNomVar07}
of the subtyping relation, and

\noindent \quad{}$\bullet$\enskip{}Mutually-recursive definitions
in OOP (\emph{e.g.}, of classes, and of the subtyping and containment
relations) can be modeled by mutually-(co)inductive mathematical objects.\vspace{0.05cm}

Additionally we observe that, by incorporating nominal subtyping,
the presented order-theoretic model of generics crucially depends
on the \emph{finite} inheritance relation between classes\footnote{Since it is always explicitly declared using class \emph{names}, inherit\-ance/sub\-classing
is an inherently nominal relation.}. On the other hand, extant models of generic OOP---which capture
conversion and bounded existentials are prominent characteristics
of---are inspired by structural (\emph{i.e.}, non-nominal) models
of polymorphic functional programming. Those models thus largely ignore
the nominal subclassing relation---explicitly declared by OO software
developers---when interpreting the generic subtyping relation and
other features of generic OOP. Influenced by their origins in functional
programming, those models depend instead on concepts and tools\footnote{Such as existentials, abstract datatypes, and the opening/closing
of type ``packages.''} developed for structural typing and structural subtyping.

On account of these observations we believe the order-theoretic model
of generics is a significantly simpler and more intuitive model of
generic OOP than extant models, and that it is more in the spirit
of nominally-typed OO type systems than those models are.

\bibliographystyle{plain}

\end{document}